\documentclass[%
 reprint,
preprintnumbers,
 amsmath,amssymb,
 aps,
prb,
uplatex
]{revtex4-2}

\usepackage{graphicx}
\usepackage{dcolumn}
\usepackage{bm}

\usepackage{braket}
\usepackage{longtable}
\usepackage{txfonts}
\usepackage{ulem}
\usepackage{multirow}

\begin{document}

\preprint{APS/123-QED}

\title{
Analysis of model parameter dependences on the second-order nonlinear conductivity in $\mathcal{PT}$-symmetric collinear antiferromagnetic metals with magnetic toroidal moment on zigzag chains 
}

\author{Megumi Yatsushiro$^{1,3}$, Rikuto Oiwa$^{2}$, Hiroaki Kusunose$^{2}$, and Satoru Hayami$^{3}$}
 \affiliation{$^1$Department of Physics, Hokkaido University, Sapporo 060-0810, Japan \\
 $^2$Department of Physics, Meiji University, Kawasaki 214-8571, Japan \\
 $^3$ Department of Applied Physics, The University of Tokyo, Tokyo 113-8656, Japan}

\begin{abstract}
A magnetic toroidal moment is a fundamental electronic degree of freedom in the absence of both spatial inversion and time-reversal symmetries and gives rise to novel multiferroic and transport properties.
We elucidate essential model parameters of the nonlinear transport in the space-time ($\mathcal{PT}$) symmetric collinear antiferromagnetic metals accompanying a magnetic toroidal moment.
By analyzing the longitudinal and transverse components of the second-order nonlinear conductivity on a two-dimensionally stacked zigzag chain based on the nonlinear Kubo formula, we show that an effective coupling between the magnetic toroidal moment and the antisymmetric spin-orbit interaction is an essential source of the nonlinear conductivity.
Moreover, we find that the nonreciprocal longitudinal current and nonlinear transverse current in a multi-band system are largely enhanced just below the transition temperature of the antiferromagnetic ordering.
We also discuss the relevance of the nonlinear conductivity to the linear magnetoelectric coefficient and conductivity.
Our result serves as a guide for exploring microscopic essence  and clarifying the parameter dependence of the nonlinear conductive phenomena in ferrotoroidal metals.
\end{abstract}

\maketitle

\section{Introduction \label{sec:introduction}}

Spontaneous time-reversal symmetry breaking has long been attracted much attention,
as it leads to intriguing physical phenomena, such as the anomalous Hall effect and the magneto-optical Kerr effect.
Modern understanding of these phenomena has been achieved based on the Berry phase mechanism~\cite{Nagaosa_review,xiao2010berry}.
Although such phenomena were originally studied in the ferromagnetic state,
it has recently been recognized that similar phenomena can occur in a certain class of antiferromagnetic (AFM) states without the uniform magnetization~\cite{smejkal2021anomalous}.
For example, the collinear AFM ordering with the mirror symmetry breaking as the uniform magnetization, results in the anomalous Hall effect~\cite{PhysRevB.55.8060, li2019quantum, PhysRevB.102.075112, PhysRevB.103.L180407}.
Thus, the AFM materials can also exhibit the same physical properties as ordinary ferromagnetic ones, which is advantageous for functional materials without leakage of a magnetic field.

The AFM state also exhibits multiferroic phenomena when both spatial inversion $(\mathcal{P})$ and time-reversal $(\mathcal{T})$ symmetries are broken simultaneously while their product $(\mathcal{PT})$ symmetry is preserved.
The typical example is the linear magnetoelectric effect in the AFM insulators, e.g., Cr$_{2}$O$_{3}$~\cite{popov1999magnetic}, Ga$_{2-x}$Fe$_x$O$_3$~\cite{popov1998magnetoelectric, doi:10.1143/JPSJ.74.1419}, LiCoPO$_4$~\cite{van2007observation, zimmermann2014ferroic}, and Ba$_2$CoGe$_2$O$_7$~\cite{PhysRevB.84.094421}, and in the AFM metals, e.g., UNi$_{4}$B~\cite{hayami2014toroidal, hayami2015toroidal, saito2018evidence} and Ce$_3$TiBi$_5$~\cite{doi:10.7566/JPSCP.30.011189, doi:10.7566/JPSJ.89.033703}.
Moreover, the nonreciprocal optical and transport properties have been studied~\cite{tokura2018nonreciprocal, ma2021topology, oyamadaa2018anomalous, PhysRevResearch.2.043081, PhysRevX.11.011001}.
Among them, multiferroic phenomena {\it within} the linear response theory have been understood by regarding the fact that the AFM states accompany the uniform orderings of the electronic odd-parity magnetic-type multipoles~\cite{Yanase_JPSJ.83.014703,hayami2014toroidal,hayami2014spontaneous,hayami2016emergent,
watanabe2017magnetic,
yanagi2018manipulating, yanagi2018theory, PhysRevB.98.020407,
hayami2018classification, watanabe2018group,
PhysRevB.98.060402,PhysRevB.98.020407,thole2018magnetoelectric, yatsushiro2019atomic,
thole2020concepts, jpsj_yatsushiro2020odd,PhysRevB.104.045117, PhysRevB.104.054412},
such as the magnetic toroidal (MT) dipole
~\cite{doi:10.1080/00150199408213381, PhysRevB.76.214404, spaldin2008toroidal, Kopaev_2009,
hayami2014toroidal, hayami2014spontaneous,hayami2016emergent,
PhysRevB.97.134423, PhysRevB.98.060402,PhysRevB.98.020407,yatsushiro2019atomic,  jpsj_yatsushiro2020odd}.

Meanwhile, the microscopic understanding of the {\it nonlinear} transports in AFMs has not been fully achieved except for several works~\cite{PhysRevResearch.2.043081,PhysRevLett.127.277201, PhysRevLett.127.277202} and symmetry analyses~\cite{PhysRevB.104.054412, zhang2020higher}.
For example, it remains unclear which model parameters are essentially important to induce nonlinear transports and how the odd-parity magnetic-type multipoles are related to them.
To be clear this point and obtain an intuitive understanding of the nonlinear transport, it is useful to extract the essential model parameters, without which the nonlinear transport coefficients vanish, from various hopping processes, spin-orbit coupling, and order parameters in the microscopic model Hamiltonian.
Such an understanding provides a guideline to
explore new functional AFM materials with a giant nonlinear transport, and its efficient bottom-up design in combination with the {\it ab initio} calculations.

In this paper, we elucidate the microscopic essential model parameters for the second-order nonlinear conductivity in the $\mathcal{PT}$-symmetric collinear AFMs by focusing on the role of the MT moment.
By analyzing a minimal model on a two-dimensionally-stacked zigzag chain based on the nonlinear Kubo formula,
we show that the effective coupling between the MT moment and one of the antisymmetric spin-orbit interactions (ASOIs) plays an essential role in inducing the longitudinal and transverse components of the nonlinear conductivity.
Moreover, we find that the nonlinear conductivities are highly enhanced near the transition temperature in the case that the AFM molecular field is comparable to the ASOI in a multi-band system.
We also discuss the relevance between the transverse nonlinear conductivity and the linear magnetoelectric coefficient by comparing the ASOI and temperature dependences.

The organization of this paper is as follows.
In Sec.~\ref{sec:model}, we introduce a minimal model on a two-dimensionally stacked zigzag chain.
After showing the relation of an MT moment to the nonlinear conductivity and the linear magnetoelectric coefficient in Sec.~\ref{sec:formalism}, the numerical results are presented in Sec.~\ref{sec:numerical}.
In Sec.~\ref{sec:discussion}, we discuss the essential model parameters and the semi-quantitative evaluation of the nonlinear conductivity.           
We summarize this paper in Sec.~\ref{sec:summary}.
In Appendix~\ref{sec:MP_express},
we present the functional forms of the odd-parity magnetic and MT multipoles.
In Appendix~\ref{sec:extract_parameter}, we show the analytic expressions for the essential model parameters in the asymmetric band modulation, nonlinear conductivities, and linear magnetoelectric coefficient.
Finally, we present the numerical result of the nonlinear transverse conductivity in the presence of the additional interlayer hopping in Appendix~\ref{sec:interlayer_hopping}.

\begin{figure}[t!]
\centering
\includegraphics[width=88mm]{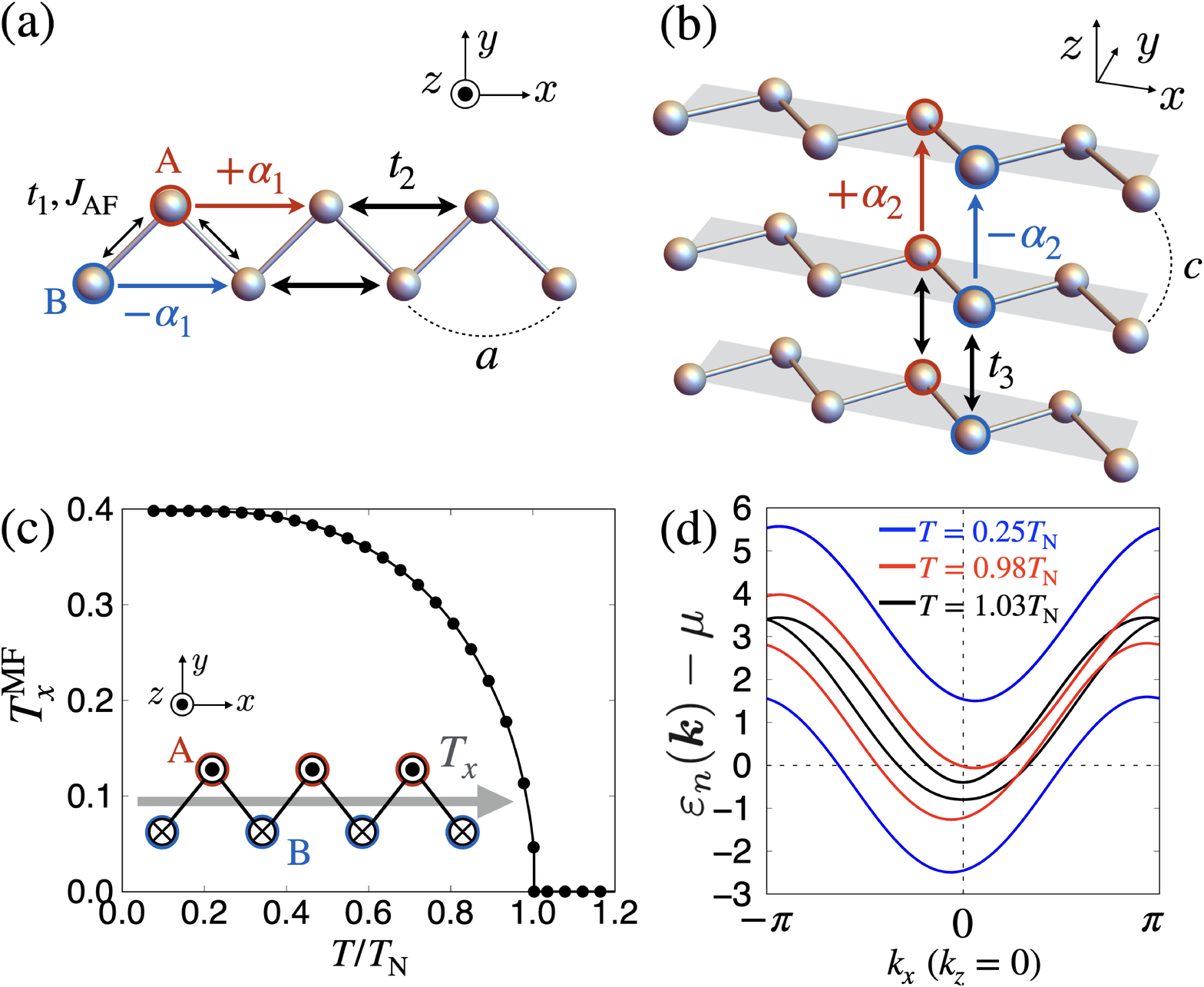}
\caption{(a), (b) Schematic pictures of (a) a two-sublattice zigzag chain and (b) its stacking along the $z$ direction.
(c) The temperature ($T$) dependence of the MT moment $T_x^{\rm MF}$ at $\alpha_1=0.4$ and $\alpha_2=0.1$.
The AFM structure with the MT moment along the $x$ direction $T_x$ is shown in the inset.
(d) The energy bands measured from the chemical potential $\mu$ at $k_z=0$ for three temperatures.
 \label{fig:fig1}}
\end{figure}

\section{Model \label{sec:model}}

We consider a minimal two-dimensional system where the zigzag chain along the $x$ direction [Fig.~\ref{fig:fig1}(a)] is stacked along the $z$ direction [Fig.~\ref{fig:fig1}(b)].
The tight-binding Hamiltonian is given by
\begin{align}
\label{eq:hamiltonian}
\mathcal{H}=& \mathcal{H}_{\rm hop}^{\rm AB} + \mathcal{H}_{\rm hop} + \mathcal{H}_{\rm ASOI} + \mathcal{H}_{\rm int},\\
\label{eq:hop_different}
\mathcal{H}_{\rm hop}^{\rm AB} =& \sum_{\bm k}  \sum_{\sigma}
\left\{\varepsilon^{\rm AB}({\bm k}) c_{{\bm k}{\rm A}\sigma}^\dagger c_{{\bm k}{\rm B}\sigma}
+ {\rm H.c.}
\right\},\\
\label{eq:hop_same}
\mathcal{H}_{\rm hop} =& \sum_{\bm k}  \sum_{\sigma}
\varepsilon({\bm k})(c_{{\bm k}{\rm A}\sigma}^\dagger c_{{\bm k}{\rm A}\sigma}+c_{{\bm k}{\rm B}\sigma}^\dagger c_{{\bm k}{\rm B}\sigma}), \\
\label{eq:ASOI}
\mathcal{H}_{\rm ASOI} =& \sum_{\bm k} \sum_{\sigma\sigma'}
{\bm g}({\bm k}) \cdot {\bm \sigma}^{\sigma\sigma'}(c_{{\bm k}{\rm A}\sigma}^\dagger c_{{\bm k}{\rm A}\sigma'}-c_{{\bm k}{\rm B}\sigma}^\dagger c_{{\bm k}{\rm B}\sigma'}), \\
\label{eq:int}
\mathcal{H}_{\rm int}=&
J_{\rm AF}
\sum_{\braket{ij}}
\hat{M}_{i{\rm A}}^{z} \hat{M}_{j{\rm B}}^{z},
\end{align}
where $c_{{\bm k} l \sigma}^\dagger$ ($c_{{\bm k} l \sigma}$) is the creation (annihilation) operator of
electrons at wave vector ${\bm k}$, sublattice $l= {\rm A}, {\rm B}$, and spin $\sigma = \uparrow, \downarrow$.
The hopping Hamiltonian $\mathcal{H}_{\rm hop}^{\rm AB}$ in Eq.~(\ref{eq:hop_different}) includes the nearest-neighbor hopping between A and B sublattices as $\varepsilon^{\rm AB}({\bm k})=-2t_1\cos (k_xa/2)$, while $\mathcal{H}_{\rm hop}$ includes the hoppings within the same sublattices along the $x$ and $z$ directions as $\varepsilon({\bm k})= -2t_2\cos{(k_xa)} -2t_3\cos{(k_zc)}$.
$\mathcal{H}_{\rm ASOI}$ in Eq.~(\ref{eq:ASOI}) represents the ASOI that arises from the relativistic spin-orbit coupling as ${\bm g}({\bm k}) = [-\alpha_2 \sin{(k_zc)}, 0, \alpha_1 \sin{(k_xa)}]$.
The ASOI in Eq.~\eqref{eq:ASOI} has the sublattice-dependent staggered form satisfying the global inversion symmetry~\cite{doi:10.1143/JPSJ.81.034702, Yanase_JPSJ.83.014703}.
$\mathcal{H}_{\rm int}$ in Eq.~\eqref{eq:int} represents the Ising-type AFM exchange interaction of the nearest-neighbor A-B bond with $J_{\rm AF} > 0$ where $\hat{M}_{i{\rm A}({\rm B})}^z = \sum_{\sigma\sigma'} c_{i{\rm A}({\rm B})\sigma}^\dagger \sigma^z_{\sigma\sigma'} c_{i{\rm A}({\rm B})\sigma'}$ is the $z$ component of the magnetic dipole operator and $c_{il\sigma}^\dagger$ and $c_{il\sigma}$ are the Fourier transforms of $c_{\bm{k}l\sigma}^\dagger$ and $c_{\bm{k}l\sigma}$, respectively.
We adopt the Hartree-type mean-field approximation as
\begin{align}
&J_{\rm AF}\sum_{\braket{ij}} \hat{M}_{i{\rm A}}^z \hat{M}_{j{\rm B}}^z \notag\\
&\to
\tilde{J}_{\rm AF}\sum_i \left(
 \braket{\hat{M}_{\rm A}^z} \hat{M}_{i{\rm B}}^z +  \braket{\hat{M}_{\rm B}^z} \hat{M}_{i{\rm A}}^z  -  \braket{\hat{M}_{\rm A}^z}\braket{\hat{M}_{\rm B}^z}\right),
 \end{align}
 where $\braket{\cdots}$ represents the statistical average and $\tilde{J}_{\rm AF}=2J_{\rm AF}$ is the renormalized coupling constant taking into account the two nearest-neighbor atomic sites.
We set the model parameters as $(t_1, t_2, t_3, J_{\rm AF})=(0.1, 1, 0.5, 2.5)$, electron filling as $1/5$, and the lattice constant as $a=c=1$ in the following discussion; $t_2$ is set as the energy unit.

The model in Eq.~(\ref{eq:hamiltonian}) exhibits the MT moment when the global inversion symmetry is broken under the staggered AFM ordering,
as shown in the inset of Fig.~\ref{fig:fig1}(c)~\cite{Yanase_JPSJ.83.014703,hayami2014toroidal}.
In the present system, the staggered AFM moment along the $z$ direction is equivalent to the uniform MT moment along the $x$ direction; $T_x^{\rm MF} \equiv (\braket{\hat{M}_{\rm A}^z} - \braket{\hat{M}_{\rm B}^z})/2$~\cite{hayami2015spontaneous}; see also Appendix~\ref{sec:MP_express}.
The $T$ dependence of $T_x^{\rm MF}$ at $\alpha_1=0.4$ and $\alpha_2=0.1$ is shown in Fig.~\ref{fig:fig1}(c), where
$T_x^{\rm MF}$ is self-consistently determined for the two-sublattice unit cell by taking over $200^2$ grid points in the Brillouin zone.
$T_x^{\rm MF}$ becomes nonzero below the transition temperature $T_{\rm N}$ and saturates below $T \simeq 0.2 T_{\rm N}$.
Almost the same behavior is obtained for $\alpha_1,\alpha_2 \lesssim 0.5$.
Reflecting $T_x^{\rm MF} \neq 0$, the electronic band structure is asymmetrically modulated along the $k_x$ direction, as shown in Fig.~\ref{fig:fig1}(d)~\cite{Yanase_JPSJ.83.014703,hayami2015spontaneous}.
This asymmetric band modulation is understood from the effective coupling between
$T_x^{\rm MF}$ and the ASOI $\alpha_1$ in the doubly degenerate bands with the $\mathcal{PT}$ symmetry, i.e.,
\begin{align}
\label{eq:E}
\varepsilon_{\pm}({\bm k})&=\varepsilon ({\bm k}) \pm X({\bm k}),\\
X({\bm k})&=\sqrt{(\alpha_1 s_x- \tilde{T}_x^{\rm MF})^2 + \alpha^2_2 s^2_z  + 4t^2_1c^2_{x/2}},
\end{align}
where $s_x=\sin k_x$, $s_z=\sin k_z$,
$c_{x/2}=\cos k_x/2$, and $\tilde{T}_x^{\rm MF} =\tilde{J}_{\rm AF}T_x^{\rm MF}$.
The factor $(\alpha_1 s_x- \tilde{T}_x^{\rm MF})^2$ includes the coupling between $\tilde{T}_x^{\rm MF}$ and $\alpha_1$ with the odd function of $k_x$.
This asymmetric band modulation due to the coupling between $\alpha_1$ and  $\tilde{T}_x^{\rm MF}$ becomes a source of the nonlinear transport as will be discussed in the following sections; see also Appendix~\ref{sec:extract_parameter}.

\section{Second-order nonlinear conductivity and linear response coefficient \label{sec:formalism}}

\subsection{Second-order nonlinear conductivity \label{sec:formalism_nonlinear}}
The second-order nonlinear conductivity tensor $\sigma_{\mu\nu\lambda}$ defined as $J_{\mu}=\sigma_{\mu\nu\lambda}E_{\nu} E_{\lambda}$ ($\mu,\nu,\lambda=x,y,z$) is calculated on the basis of the second-order Kubo formula~\cite{PhysRevResearch.2.043081}.
In the clean limit, the intraband contribution is dominant, which is given by
\begin{align}
\label{eq:ncon}
\sigma_{\mu\nu\lambda} =
\frac{e^3\tau^2}{\hbar^3}\frac{1}{V} \sum_{\bm k} \sum_{n} \frac{\partial^2 \varepsilon_n({\bm k})}{\partial k_\mu\partial k_\nu }
\frac{\partial \varepsilon_n({\bm k})}{\partial k_\lambda} \frac{\partial f [\varepsilon_n ({\bm k})]}{\partial \varepsilon_n({\bm k})},
\end{align}
where $e(>0)$, $\tau$, $\hbar$, and $V$ are the elementary charge, relaxation time, the reduced Planck constant, and the system volume, respectively~\footnote{There is no contribution from the Berry curvature dipole~\cite{sodemann2015} because of the $\mathcal{PT}$ symmetry, while the interband contribution in the $\mathcal{T}$-breaking system~\cite{gao2014} is neglected by considering the clean limit.}.
$f[\varepsilon_n({\bm k})]$ is the Fermi distribution function for the $n$th-band eigen energy $\varepsilon_n({\bm k})$.
The intraband contribution in Eq.~\eqref{eq:ncon} represents the Drude-type one with the dissipation $\tau^{-2}$, whose expression eventually coincides with that obtained by the Boltzmann formalism~\cite{ideue2017bulk, PhysRevB.99.045121, PhysRevResearch.2.043081, gao2019semiclassical}.
Hereafter, we use the scaled $\sigma_{\mu\nu\lambda}$ as $\bar{\sigma}_{\mu\nu\lambda} =\sigma_{\mu\nu\lambda} /(e^3\tau^2\hbar^{-3})$.

From Eq.~(\ref{eq:ncon}), one finds the relation $\sigma_{\mu\nu\nu}=\sigma_{\nu\mu\nu}$ by integration by parts.
This indicates that the Drude-type nonlinear conductivity $\sigma_{\mu\nu\lambda}$ is the totally symmetric rank-3 tensor with 10 independent components: $\sigma_{xxx}$, $\sigma_{yyy}$, $\sigma_{zzz}$, $\sigma_{xyy}$, $\sigma_{yzz}$, $\sigma_{zxx}$, $\sigma_{xxy}$, $\sigma_{yyz}$, $\sigma_{zzx}$, and $\sigma_{xyz}$.
As $\sigma_{\mu\nu\lambda}$ is a third-rank polar time-reversal-odd tensor, i.e., $ \sigma_{\mu\nu\lambda} \to -\sigma_{\mu\nu\lambda}$ under $\mathcal{P}$ or $\mathcal{T}$ operation but $\sigma_{\mu\nu\lambda} \to \sigma_{\mu\nu\lambda}$ under $\mathcal{PT}$ operation, it becomes nonzero when both the spatial inversion and time-reversal symmetries are absent.
From the multipole viewpoint, above symmetry requirement means that the nonzero tensor components are related to the active odd-parity MT multipoles~\cite{dubovik1974multipole, dubovik1986axial, dubovik1990toroid, spaldin2008toroidal, hayami2018microscopic}: three rank-1 MT dipoles $(T_x, T_y, T_z)$ and seven rank-3 MT octupoles $(T_{xyz}, T_x^\alpha, T_y^\alpha, T_z^\alpha, T_x^\beta, T_y^\beta, T_z^\beta)$, whose correspondence is given by~\cite{PhysRevB.104.054412}
 \begin{align}
\label{eq:rank12}
\sigma &=
\begin{pmatrix}
\sigma_{xxx} & \sigma_{yxx} & \sigma_{zxx} \\
\sigma_{xyy} & \sigma_{yyy} & \sigma_{zyy} \\
\sigma_{xzz} & \sigma_{yzz} & \sigma_{zzz} \\
\sigma_{xyz} & \sigma_{yyz} & \sigma_{zyz} \\
\sigma_{xzx} & \sigma_{yzx} & \sigma_{zzx} \\
\sigma_{xxy} & \sigma_{yxy} & \sigma_{zxy} \\
\end{pmatrix}^{\rm T}
\notag\\
&\leftrightarrow
\begin{pmatrix}
3{T}_x + 2T_{x}^\alpha
 &T_y - T_{y}^\alpha -T_{y}^\beta
 & T_z - T_{z}^\alpha + T_{z}^\beta \\
T_x - T_{x}^\alpha + T_{x}^\beta
 &3{T}_y + 2T_{y}^\alpha
 &T_z - T_{z}^\alpha - T_{z}^\beta \\
T_x - T_{x}^\alpha -T_{x}^\beta
 & T_y -T_{y}^\alpha + T_{y}^\beta
 &3{T}_z + 2T_{z}^\alpha \\
 T_{xyz}
 &T_z - T_{z}^\alpha - T_{z}^\beta
 & {T}_y - T_{y}^\alpha + T_{y}^\beta \\
{T}_z - T_{z}^\alpha + T_{z}^\beta
 & T_{xyz}
 & {T}_x -T_{x}^\alpha - T_{x}^\beta \\
{T}_y - T_{y}^\alpha - T_{y}^\beta
 & {T}_x - T_{x}^\alpha + T_{x}^\beta
  & T_{xyz}\\
\end{pmatrix}^{\rm T},
\end{align}
where the functional forms of dipoles and octupoles are summarized in Appendix~\ref{sec:MP_express}.
The correspondence in Eq.~\eqref{eq:rank12} is obtained by decomposing $\sigma_{\mu\nu\lambda}$ into the tensor components with the same rotational symmetry to the dipoles and octupoles (See also Appendix~\ref{sec:MP_express}).
When the MT dipole and/or MT octupole in Eq.~\eqref{eq:rank12} are activated in an AFM metal, the corresponding tensor component of $\sigma_{\mu\nu\lambda}$ becomes nonzero.
From Eq.~(\ref{eq:rank12}), one finds that MT dipole $T_\mu$ is relevant to the longitudinal component $\sigma_{\mu\mu\mu}$ and the transverse components $\sigma_{\mu\nu\nu}$ and $\sigma_{\nu\mu\nu}$ ($\nu\neq\mu$).
It means that both nonreciprocal conductivity and nonlinear transverse conductivity are expected to be realized in the presence of the MT dipole, i.e., ferrotoroidal metals~\cite{litvin2008ferroic,schmid2008some,PhysRevB.104.054412}.

In the present system under the magnetic point-group $m'mm$ with the nonzero MT moment $T_x^{\rm MF}$,
five components $\sigma_{xxx}$, $\sigma_{xyy}$, $\sigma_{yxy}$, $\sigma_{xzz}$, and $\sigma_{zzx}$ can be nonzero, since $T_x$, $T_x^\alpha$, and $T_x^\beta$ in Eq.~\eqref{eq:rank12} belong to the totally symmetric irreducible representation~\cite{PhysRevB.104.054412}.
Among them, $\sigma_{xyy}$ and $\sigma_{yxy}$ vanish owing to $k_y=0$ in the present two-dimensional system.
In addition to the nonzero contribution from the linear conductivity $\sigma_{xx}$,
$\sigma_{xxx}$  results in the nonreciprocal current, while  $\sigma_{xzz}$ without linear $\sigma_{xz}$ leads to the pure second-order transverse current, respectively.

\subsection{Linear response coefficient \label{sec:formalism_linear}} 
In the presence of the MT moment $T_x^{\rm MF}$, the linear magnetoelectric tensor $\alpha_{\mu\nu}$ in $M_\mu=\alpha_{\mu\nu}E_{\nu}$ ($\mu,\nu=x,y,z$) is also finite.
We calculate the linear magnetoelectric tensor by the linear response theory as
\begin{align}
\label{eq:a_munu}
\alpha_{\mu\nu} =& \frac{eg\mu_{\rm B} \hbar}{2Vi} \sum_{\bm k} \sum_{n \neq m} \frac{ f[\varepsilon_n ({\bm k})]-f[\varepsilon_m({\bm k})]}{[\varepsilon_n({\bm k})-\varepsilon_m({\bm k})]^2 + (\hbar \delta)^2} \sigma_{\mu{\bm k}}^{nm} \varv_{\nu{\bm k}}^{mn},
\end{align}
where $g$ and $\mu_{\rm B}$ are the g factor ($g=2$) and Bohr magneton, respectively.
$\sigma_{\mu{\bm k}}^{nm}=\braket{n{\bm k}|\sigma_{\mu}|m{\bm k}}$ and $\varv_{\nu{\bm k}}^{mn}=\braket{m{\bm k}|\varv_{\nu{\bm k}}|n{\bm k}}$ are the matrix elements of spin $\sigma_{\mu}$ and velocity $\varv_{\nu{\bm k}}=\partial \mathcal{H}/(\hbar \partial k_{\nu})$ for the eigenstate $\ket{n{\bm k}}$.
We use the scaled $\bar{\alpha}_{\mu\nu} = \alpha_{\mu\nu}/(e\mu_{\rm B}\hbar)$ in the following discussion.

As $\alpha_{\mu\nu}$ in a $\mathcal{PT}$ symmetric system is relevant to the rank-0--2 odd-parity multipoles: magnetic monopole $M_0$, MT dipoles $(T_x, T_y, T_z)$, and magnetic quadrupoles $(M_u, M_v, M_{yz}, M_{zx}, M_{xy})$ (see also Appendix~\ref{sec:MP_express}), the relation is represented as follows~\cite{hayami2018classification, watanabe2018group}:
\begin{align}
\alpha &=
\begin{pmatrix}
\alpha_{xx} & \alpha_{xy} & \alpha_{xz}\\
\alpha_{yx} & \alpha_{yy} & \alpha_{yz}\\
\alpha_{zx} & \alpha_{zy} & \alpha_{zz}\\
\end{pmatrix} \\
&\leftrightarrow
\begin{pmatrix}
M_0 -M_u+M_v & M_{xy}+T_z & M_{zx}-T_y\\
M_{xy}-T_z & M_0-M_u-M_v & M_{yz}+T_x \\
M_{zx}+T_y & M_{yz}-T_x & M_0+2M_u \\
\end{pmatrix}.
\label{eq:formalism_ME_mul}
\end{align}
Since $T_x$ and $M_{yz}$ become active for $T_x^{\rm AF}\neq 0$ in the present system,
$\alpha_{yz}$ and $\alpha_{zy}$ are expected to be nonzero.
As $\alpha_{zy}$ is zero due to the two dimensionality, we only consider $\alpha_{yz}$.

For the following discussion, we also present the linear Hall conductivity
\begin{align}
\label{eq:s_xz}
\sigma_{xz} =& \frac{e^2\hbar}{Vi} \sum_{\bm k} \sum_{n \neq m} \frac{ f[\varepsilon_n ({\bm k})]-f[\varepsilon_m({\bm k})]}{[\varepsilon_n({\bm k})-\varepsilon_m({\bm k})]^2 + (\hbar \delta)^2} \varv_{x{\bm k}}^{nm} \varv_{z{\bm k}}^{mn}.
\end{align}
We use the scaled value $\bar{\sigma}_{xz}=\sigma_{xz}/(e^2\hbar H_y)$ in the following, where $H_y$ is the Zeeman field along the $y$ direction.

\section{Numerical Result \label{sec:numerical}}
\subsection{Longitudinal second-order conductivity $\sigma_{xxx}$ \label{sec:numerical_xxx}}

\begin{figure}[t!]
\centering
\includegraphics[width=85mm]{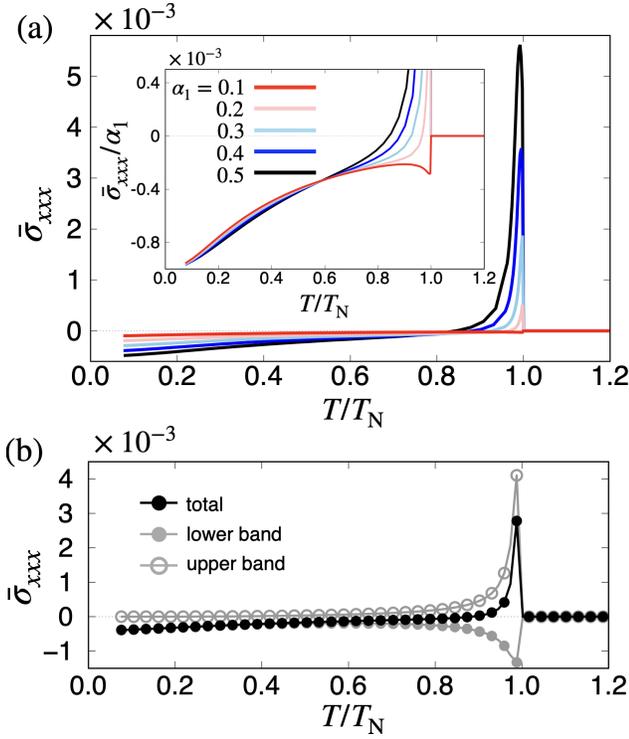}
\caption{
(a) The longitudinal second-order conductivity $\bar{\sigma}_{xxx}$ for $\alpha_1=0.1$--$0.5$ as a function of $T$ at $\alpha_2=0.1$.
The inset shows $\bar{\sigma}_{xxx}/\alpha_1$.
(b) The upper- and lower-band contributions to $\bar{\sigma}_{xxx}$ at $\alpha_1=0.4$.
\label{fig:fig2}}
\end{figure}

We first show the numerical result of the longitudinal nonlinear conductivity $\bar{\sigma}_{xxx}$.
Figure~\ref{fig:fig2}(a) shows $\bar{\sigma}_{xxx}$ as a function of $T$ for various $\alpha_1=0.1$--$0.5$ at $\alpha_2=0.1$.
The $T$ dependence for different $\alpha_1$ is qualitatively similar; $\bar{\sigma}_{xxx}$  is largely enhanced just below $T=T_{{\rm N}}$, and shows maximum with decrease of $T$.
While further decreasing $T$, $\bar{\sigma}_{xxx}$ shows the sign change, and then reaches a negative value at the lowest $T$.

The nonzero $\sigma_{xxx}$ is closely related to the formation of the asymmetric band structure under $T_x^{\rm MF} \neq 0$, since $\sigma_{xxx}$ has the same symmetry as $T_x^{\rm MF}$~\cite{PhysRevB.104.054412}.
As the asymmetric band modulation is caused by the coupling between $\tilde{T}_x^{\rm MF}$ and $\alpha_1$,
they are indispensable for nonzero $\sigma_{xxx}$.
Indeed, $\bar{\sigma}_{xxx}$ vanishes for $\alpha_1=0$ or $\tilde{T}_x^{\rm MF}=0$.
Moreover, $\bar{\sigma}_{xxx}$ is well scaled by $\bar{\sigma}_{xxx}/\alpha_1$ at low temperatures $T \lesssim 0.7T_{\rm N}$ for small $\alpha_1$.
See Sec.~\ref{sec:Essential model parameters} for the essential model parameters in details.

Meanwhile, $\bar{\sigma}_{xxx}$ is not scaled by $\alpha_1$ for $0.7\lesssim T/T_{\rm N}\leq 1$ in the region where $\bar{\sigma}_{xxx}$ is drastically enhanced.
This is attributed to the rapid increase of $\tilde{T}_x^{\rm MF}$ and resultant drastic change of the electronic band structure near the Fermi level.
As $\bar{\sigma}_{xxx}$ in Eq.~(\ref{eq:ncon}) includes the factors $\partial^2 \varepsilon_n({\bm k})/\partial k_x^2 $ and $\partial \varepsilon_n({\bm k})/\partial k_x$, the small $X(\bm{k})$ appearing in the denominator of $\partial^2 \varepsilon_n({\bm k})/\partial k_x^2 $ and $\partial \varepsilon_n({\bm k})/\partial k_x$
gives a dominant contribution.
When considering the small order parameter compared to the ASOI, i.e., $\tilde{T}_x^{\rm MF} \lesssim \alpha_1$, $X(\bm{k})$ can become small when the Fermi wavenumber $k^{\rm F}_x$ satisfies $\tilde{T}_x^{\rm MF}  \simeq \alpha_1 \sin k^{\rm F}_x$, which results in a large enhancement of $\bar{\sigma}_{xxx}$.
Such an enhancement is remarkable when the upper and lower bands are closely located in the paramagnetic state with small $X(\bm{k})$ as shown in Fig.~\ref{fig:fig1}(d), which can be realized for small $t_1=0.1$ and $\alpha_2=0.1$.
In short, there are two conditions for large $\bar{\sigma}_{xxx}$:
One is the large essential parameters, such as $\alpha_1$, $T_x^{\rm MF}$, and $J_{\rm AF}$,
and the other is to satisfy $\tilde{T}_x^{\rm MF}  \simeq \alpha_1 \sin k^{\rm F}_x$ in a multi-band system.
These conditions can be experimentally controlled by electron/hole doping and temperature.

The sign change of $\bar{\sigma}_{xxx}$ in $T$ dependence is owing to the multiband effect.
As shown in Fig.~\ref{fig:fig1}(d), the band bottom is shifted in the opposite direction for the upper and lower bands, which means that the opposite sign of the coupling $\alpha_1 \tilde{T}_x^{\rm MF}$ results in the opposite contribution to $\bar{\sigma}_{xxx}$.
This is demonstrated by decomposing $\bar{\sigma}_{xxx}$ into the upper- and lower-band contributions, as shown in Fig.~\ref{fig:fig2}(b).
The results indicate that the dominant contribution of $\bar{\sigma}_{xxx}$ arises from the upper band for $0.9\lesssim T/T_{\rm N}\leq 1$, while that arises from the lower band for $T/T_{\rm N}\lesssim 0.9$.
The suppression of the upper-band contribution for low $T$ is because it becomes away from the Fermi level by the development of $T_x^{\rm MF}$.

\subsection{Transverse second-order conductivity $\sigma_{xzz}$ \label{sec:numerical_xzz}}

\begin{figure}[htb!]
\centering
\includegraphics[width=90mm]{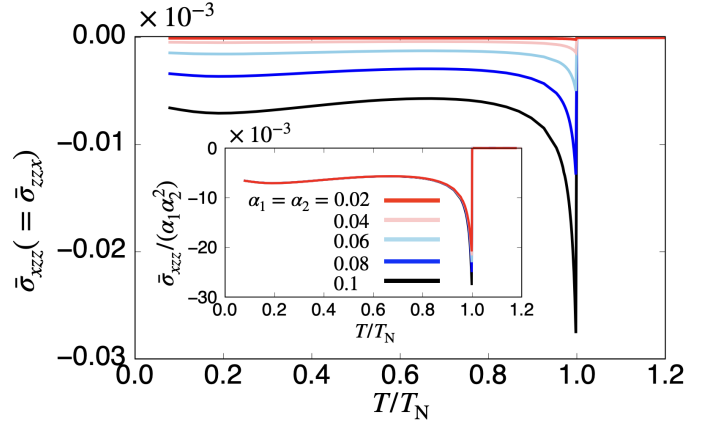}
\caption{
The transverse second-order nonlinear conductivity $\bar{\sigma}_{xzz}$ for several $\alpha_1$ and $\alpha_2$ with $\alpha_1=\alpha_2$.
The inset represents $\bar{\sigma}_{zxx}/(\alpha_1\alpha_2^2)$.
\label{fig:fig3}}
\end{figure}

Next, let us discuss the transverse nonlinear conductivity $\bar{\sigma}_{xzz}$.
Figure~\ref{fig:fig3} shows the $T$ dependence of $\bar{\sigma}_{xzz}$ for $0.02 \leq \alpha_1, \alpha_2 \leq 0.1$ with $\alpha_1=\alpha_2$.
The behavior of $\bar{\sigma}_{xzz}$ against $T$ is similar to $\bar{\sigma}_{xxx}$ except for the sign change; $\bar{\sigma}_{xzz}$ becomes nonzero below $T=T_{\rm N}$ and shows the maximum near $T_{\rm N}$.
While decreasing $T$, $\bar{\sigma}_{xzz}$ is suppressed and shows an almost constant value.

Similar to $\sigma_{xxx}$, the origin of nonzero
$\sigma_{xzz}$ is the asymmetric band modulation under $T_x^{\rm MF} \neq 0$ via the effective coupling $\tilde{T}_x^{\rm MF}\alpha_1$.
Besides, we find another contribution from $\alpha_2$ for nonzero $\sigma_{xzz}$ in contrast to $\sigma_{xxx}$, where
$\bar{\sigma}_{xzz}$ is well scaled by $\alpha_1 \alpha_2^2$ as shown in the inset of Fig.~\ref{fig:fig3}, as discussed in Sec.~\ref{sec:Essential model parameters}.
The additional parameter dependence for $\alpha_2^2$ is owing to an additional symmetry between $k_z$ and $k_z+\pi$ for $\alpha_2=0$, which gives the opposite-sign contribution to $\sigma_{xzz}$ so that totally $\sigma_{xzz}=0$.

\subsection{Comparison to magnetoelectric coefficient $\alpha_{yz}$ \label{sec:numerical_yz}}

\begin{figure}[htb!]
\centering
\includegraphics[width=90mm]{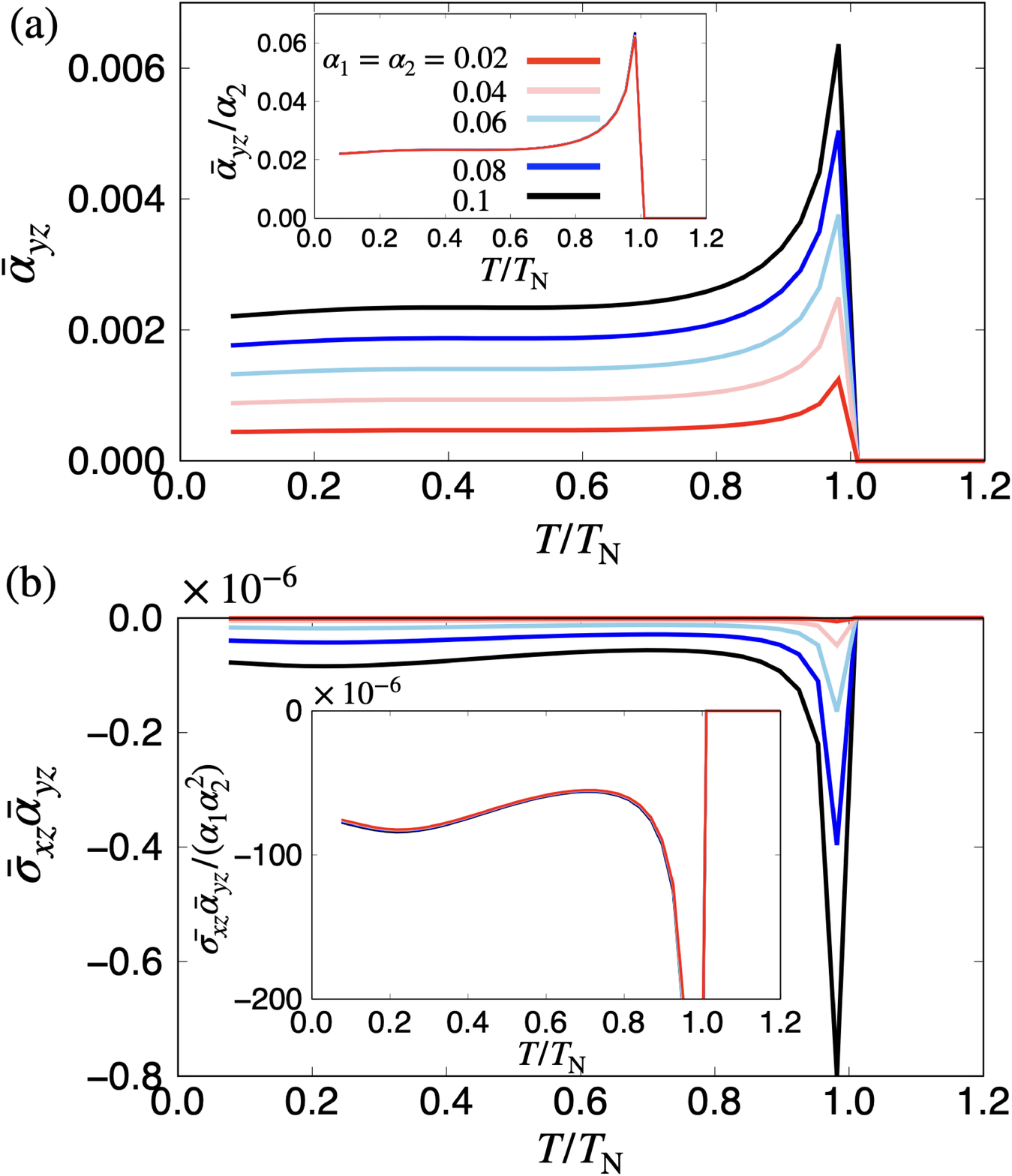}
\caption{
(a) The magnetoelectric coefficient $\bar{\alpha}_{yz}$ and
(b) the quantity $\bar{\sigma}_{xz}\bar{\alpha}_{yz}$ with the same parameters as Fig.~\ref{fig:fig3}.
$\bar{\sigma}_{xz}$ is calculated by supposing the magnetic field $H_y=0.01$.
The insets of (a) and (b) represent $\bar{\alpha}_{yz}/\alpha_2$ and $\bar{\sigma}_{xz} \bar{\alpha}_{yz}/(\alpha_1\alpha_2^2)$, respectively.
\label{fig:fig4}}
\end{figure}

We also present another MT-moment-driven phenomena, the magnetoelectric response, and compare its parameter and $T$ dependence to the nonlinear conductivities obtained in the previous section.
Figure~\ref{fig:fig4}(a) shows the $T$ dependence of $\bar{\alpha}_{yz}$ for $0.02 \leq \alpha_1, \alpha_2 \leq 0.1$ with $\alpha_1=\alpha_2$, whose behavior is similar to the transverse nonlinear conductivity $\sigma_{xzz}$ in Fig.~\ref{fig:fig3} except for the sign.
$\bar{\alpha}_{yz}$ is nonzero even if $\alpha_1=0$ that is different from the nonlinear conductivities, whereas $\alpha_2$ and $\tilde{T}_x^{\rm MF}$
are essential to obtain the finite $\bar{\alpha}_{yz}$, as detailed in Sec.~\ref{sec:Essential model parameters}.
As shown in the inset of Fig.~\ref{fig:fig3}(a), $\bar{\alpha}_{yz}$ is well scaled as $\bar{\alpha}_{yz}/\alpha_2$ for small $\alpha_2$.

Moreover, it is noteworthy to comment on the relation between the transverse nonlinear conductivity and a combination of the linear magnetoelectric and Hall coefficients, since the nonlinear transverse transport in the $\mathcal{PT}$-symmetric AFMs can be understood as the Hall transport driven by the induced magnetization through the linear magnetoelectric response at the phenomenological level~\cite{hayami2014toroidal, oyamadaa2018anomalous}.

We show the $T$ dependence of $\bar{\sigma}_{xz}\bar{\alpha}_{yz}$ in Fig.~\ref{fig:fig4}(b) for the same parameters in Fig.~\ref{fig:fig3}.
The small magnetic field $H_y=0.01$ is introduced to mimic the induced magnetization in $\alpha_{yz}$.
Compared to the results in Fig.~\ref{fig:fig3} and \ref{fig:fig4}(b),
one finds the resemblance between the $T$ dependences of $\bar{\sigma}_{xzz}$ and $\bar{\sigma}_{xz}\bar{\alpha}_{yz}$, both of which are scaled by $\alpha_1 \alpha_2^2$.
A good qualitative correspondence in these responses indicates that the interpretation of dividing subsequent two linear processes for nonlinear conductivity is reasonable in the present model.
The overall quantitative difference $\bar{\sigma}_{xz}\bar{\alpha}_{yz}/\bar{\sigma}_{xzz}\sim 10^{-2}$ may be ascribed to the magnitude of the used internal magnetic field ($H_{y}=0.01$) that should be replaced by the true internal field. 
However, it is hard to estimate it quantitatively. 

\section{Discussion \label{sec:discussion}}
\subsection{Essential model parameters \label{sec:Essential model parameters}}

We discuss the parameter dependences of the asymmetric band modulation, nonlinear conductivity, and the linear magnetoelectric and Hall coefficients at the level of the microscopic model Hamiltonian.
For this purpose, we try to extract the essential parameters for each response from various hoppings, spin-orbit coupling, and internal/external field in the model Hamiltonian based on the method in Refs.~\onlinecite{hayami2020bottom,doi:10.7566/JPSJ.91.014701}.
In the following, we discuss the important model parameters in each case one by one, and the results are summarized in Table~\ref{tab_params}.
The derivation is shown in Appendix~\ref{sec:extract_parameter}.

First, the essential parameters for the asymmetric band modulation~\cite{hayami2020bottom} are given by $\tilde{T}_x^{\rm MF}\alpha_1$, as shown in Appendix~\ref{sec:abm}.
The result is consistent with the eigenvalues in Eq.~(\ref{eq:E}).

Next, the essential model parameters for $\sigma_{xxx}$~\cite{doi:10.7566/JPSJ.91.014701} (see also Appendix~\ref{sec:ncon}) are given by
\begin{align}
\sigma_{xxx}=&\alpha_1 \tilde{T}_x^{\rm MF} \left[t_1^2F(t_1,t_2,t_3,\alpha_1,\alpha_2, \tilde{T}_x^{\rm MF})\right. \notag\\
&\left. \qquad \quad+t_2F'(t_1,t_2,t_3,\alpha_1,\alpha_2, \tilde{T}_x^{\rm MF})\right],
\end{align}
where $F$ and $F'$ represent the arbitrary functions.
Note that only the even power of $\alpha_1$ and $\tilde{T}_x^{\rm MF}$ appears in $F$ and $F'$. 
Thus, one finds that the coupling of $\alpha_1$ and $\tilde{T}_x^{\rm MF}$ is always necessary to induce $\sigma_{xxx}$, which is consistent with the numerical result presented in Sec.~\ref{sec:numerical_xxx}.
Moreover, $\sigma_{xxx}$ is closely related to the asymmetric band modulation because both of them are characterized by the same essential model parameters.

Similarly, the essential model parameters of $\sigma_{xzz}$ are given by
\begin{align}
\sigma_{xzz}= \alpha_1\tilde{T}_x^{\rm MF} \left[\alpha_2^2 t_2  F(t_1,t_2,t_3,\alpha_1,\alpha_2,\tilde{T}_x^{\rm MF})\right],
\label{eq:para_xzz}
\end{align}
where the even power of $\alpha_1$, $\alpha_2$, and $\tilde{T}_x^{\rm MF}$ appears in $F$. 
Equation~\eqref{eq:para_xzz} shows that the coupling of $\alpha_1$ and $\tilde{T}_x^{\rm MF}$ is essential to induce $\sigma_{xzz}$ as similar to $\sigma_{xxx}$, which is consistent with the numerical result in Sec.~\ref{sec:numerical_xzz}.
Moreover, Eq.~\eqref{eq:para_xzz} indicates that $t_2$ and even power of $\alpha_2$ are also necessary for $\sigma_{xzz}$ in the present model in Eq.~\eqref{eq:hamiltonian}.

In a similar way, the essential model parameters to induce $\alpha_{yz}$ and $\sigma_{xz}$ are given by
\begin{align}
 \alpha_{yz} &= \alpha_{2} \tilde{T}_x^{\rm MF} \left[t_3 F(t_1,t_2,t_3,\alpha_1,\alpha_2,\tilde{T}_x^{\rm MF})\right], \\
\sigma_{xz} &= \alpha_{1} \alpha_{2} H_{y} \left[t_3 F(t_1,t_2,t_3,\alpha_1,\alpha_2, H_y,\tilde{T}_x^{\rm MF})\right].
 \end{align}
 This indicates that nonzero $\alpha_{yz}\sigma_{xz}$ needs nonzero $\alpha_1\alpha_2^2 \tilde{T}_x^{\rm MF}$, which shows a good agreement with the condition for $\sigma_{xzz}$.
The common essential model-parameter dependence in small parameter region was already confirmed in Secs.~\ref{sec:numerical_xzz} and \ref{sec:numerical_yz}.

It is noteworthy that the above approach to extract the essential model parameters can be straightforwardly applied even when introducing the other model parameters.
For example, let us consider the additional interlayer A-B hopping $t_4$ in the model Hamiltonian.
In this situation, one finds that there is no longer simple correlation between $\sigma_{xzz}$ and $\sigma_{xz}\alpha_{yz}$; the essential model parameters for the former are $\alpha_1 \tilde{T}_x^{\rm MF}$ rather than $\alpha_1\alpha_2^2 \tilde{T}_x^{\rm MF}$, while those for the latter still remains the same as $\alpha_1\alpha_2^2 \tilde{T}_x^{\rm MF}$ as discussed in Appendix~\ref{sec:extract_parameter}.
In other words,
the factor $\alpha^2_2 t_2$ in the square bracket in Eq.~(\ref{eq:para_xzz}) is not truly the essential factor. 
Indeed, the numerical results in the presence of $t_4$ give a different temperature dependence from each other, as shown in Appendix~\ref{sec:interlayer_hopping}.
Thus, the correspondence between $\sigma_{xzz}$ and $\sigma_{xz}\alpha_{yz}$ occurs depending on the hopping in the effective model, which is clarified by performing a procedure in Appendix~\ref{sec:extract_parameter}.

\begin{table}[h]
  \caption{
    Model parameters necessary for the asymmetric band modulation and response tensors indicated by the checkmark (\checkmark).
    In the last two columns, model parameters are decomposed into the essential and semi-essential parts.
     }
  \label{tab_params}
  \begin{center}
    \begin{tabular}{lcccccccc}
      \hline \hline
                                                                      & $t_2$      & $t_3$      & $\alpha_1$ & $\alpha_2$ & $\tilde{T}_x^{\rm MF}$ & $H_{y}$    & essential                                                                       &  
                                                                      semi-essential   \\ \hline
      asymmetric                                      &            &            & \checkmark &            & \checkmark             &            & $\alpha_1 \tilde{T}_x^{\rm MF}$                \\
      band modulation  & \\\hline

       $\sigma_{xxx}\,\, (t_4 = 0)$            &            &            & \checkmark &            & \checkmark             &            & $\alpha_1 \tilde{T}_x^{\rm MF}$ & $t_1^2, t_2$               \\
       $\sigma_{xxx}\,\, (t_4 \neq 0)$            &            &            & \checkmark &            & \checkmark             &            & $\alpha_1 \tilde{T}_x^{\rm MF}$ & $t_1^2, t_2, t_4$               \\
      $\sigma_{xzz}\,\, (t_4 = 0)$ & \checkmark &            & \checkmark & \checkmark & \checkmark             &            & $\alpha_1 \tilde{T}_x^{\rm MF}$ & $\alpha_2^2t_2$ \\
      $\sigma_{xzz}\,\, (t_4 \neq 0)$               &            &            & \checkmark &            & \checkmark             &            & $\alpha_1 \tilde{T}_x^{\rm MF}$ & $\alpha_2^2t_2, t_4$               \\ \hline
      $\alpha_{yz}\,\, (t_4 = 0)$ &            & \checkmark &            & \checkmark & \checkmark             &            & $\alpha_2  \tilde{T}_x^{\rm MF}$ & $t_3$           \\
     $\alpha_{yz}\,\, (t_4 \neq 0)$                &            &            &            & \checkmark & \checkmark             &            & $\alpha_2 \tilde{T}_x^{\rm MF}$ & $t_3, t_4$               \\ \hline
      $\sigma_{xz}\,\,(t_4 = 0)$             &            & \checkmark & \checkmark & \checkmark &                        & \checkmark & $\alpha_1 \alpha_2 H_y$  & $t_3$                  \\
      $\sigma_{xz}\,\, (t_4 \neq 0)$                &            &            & \checkmark & \checkmark &                        & \checkmark & $\alpha_1 \alpha_2 H_y$ & $t_3, t_4$                       \\
      \hline \hline
    \end{tabular}
  \end{center}
\end{table}

\subsection{Quantitative evaluation}

Finally, we discuss the order estimate of the nonlinear conductivity
for $\alpha_1=0.5$ and $\alpha_2=0.1$ by the ratio $\sigma_{xxx}/(\sigma_{xx})^2$ with being independent of
the relaxation time in the clean limit.
By putting the typical values as $a\sim 0.5$~[nm] and $|t_2|=0.2$~eV,
we obtain $\sigma_{xxx}/(\sigma_{xx})^2 \sim 10^{-3} \hbar a^{2} e^{-1} |t_2|^{-1}
\sim 10^{-18}$~[m$^3$ A$^{-1}$] for $T\to 0$ and
$10^{-17}$~[m$^3$ A$^{-1}$] near $T_{\rm N}$, which is comparable to the value in the 2D
nonmagnetic Rashba system under the magnetic field~\cite{ideue2017bulk}.
Further enhancement can be achieved by tuning the model parameters and electron filling.

\section{Summary \label{sec:summary}}
In summary, we investigated the microscopic essence for the second-order nonlinear conductivity in the $\mathcal{PT}$-symmetric collinear AFM with the MT moment on a two-dimensionally stacked zigzag chain by focusing on the role of the MT moment.
Based on the nonlinear Kubo formula in the clean limit, we found that the effective coupling between the ASOI and the MT moment is essential for the nonlinear conductivity.
By analyzing both the longitudinal and transverse components of the nonlinear conductivity while changing the ASOI and the temperature, we showed that their large enhancement can be achieved near the transition temperature, provided that the AFM molecular field is comparable to the ASOI in a multi-band system.
We also discussed the similarity and difference between the transverse nonlinear transport and the combined response consisting of the linear magnetoelectric and Hall coefficients.

The present result elucidates the essential model parameters for MT-related physical phenomena, such as the nonlinear conductivity and the linear magnetoelectric effect, in $\mathcal{PT}$-symmetric collinear AFMs.
The similar analysis can be applied to examine the role of the MT moment for any collinear AFMs with the MT moment in the zigzag structure, e.g.,
CeRu$_2$Al$_{10}$~\cite{Tursina:wm6046, doi:10.1143/JPSJ.80.073701}, Ce$_3$TiBi$_5$~\cite{doi:10.7566/JPSJ.89.033703,doi:10.7566/JPSCP.30.011189}, and $\alpha$-YbAl$_{1-x}$Mn$_x$B$_4$~\cite{PhysRevResearch.3.023140},
and other ferrotoroidal metals/semiconductors with locally noncentrosymmetric crystal structures, such as Mn$_2$Au~\cite{barthem2013revealing, PhysRevLett.127.277202}, $R$B$_4$ ($R=$Dy, Er)~\cite{WILL1981349, WILL197931},
CuMnAs~\cite{wadley2015antiferromagnetic,PhysRevLett.127.277201}, PrMnSbO~\cite{PhysRevB.82.100412}, NdMnAsO~\cite{PhysRevB.83.144429}, and $X_y$Fe$_{2-x}$Se$_2$ ($X=$K, Tl, Rb)~\cite{Bao_2011, Pomjakushin_2011, PhysRevLett.109.077003}, once the model Hamiltonian is given. 
The measurements of the linear magnetoelectric effect and the nonlinear conductivity for these materials are also useful to investigate their microscopic mechanisms.
Moreover, the analysis is straightforwardly extended to the AFMs with the other odd-parity magnetic-type multipole moments, such as the MT octupole, since they are characterized by the same spatial inversion and time-reversal symmetries.
Our study will stimulate a further investigation of the multiferroic and conductive phenomena in the $\mathcal{PT}$-symmetric AFM metals.


\begin{acknowledgements}
We thank Y. Motome and Y. Yanagi for the fruitful discussion.
This research was supported by JSPJ KAKENHI Grant Numbers
JP19K03752, JP19H01834, JP21H01037, and by JST PRESTO (JPMJPR20L8).
M.Y. and R.O. are supported by a JSPS research fellowship and JSPS KAKENHI (Grant No. JP20J12026
and JP20J21838).
\end{acknowledgements}

\appendix
\section{Expressions of multipoles \label{sec:MP_express}}
We show the functional form of multipoles with rank 0--3 except the normalization constant:
the rank 0 (monopole) is
\begin{align}
\label{eq:monopole}
X_0 \propto 1,
\end{align}
the rank 1 (dipole) is
\begin{align}
\label{eq:dipole}
(X_x, X_y, X_z) \propto (x,y,z),
\end{align}
the rank 2 (quadrupole) is
\begin{align}
\label{eq:quadrupole}
X_u& \propto 3z^2-r^2, 
\\
X_v &\propto x^2-y^2, 
\\
(X_{yz}, X_{zx}, X_{xy}) &\propto (yz, zx, xy),
\end{align}
the rank 3 (octupole) is
\begin{align}
\label{eq:octupole_1}
X_{xyz} &\propto xyz, \\
\label{eq:octupole_2}
(X_x^\alpha, X_y^\alpha, X_z^\alpha) &\propto \left(x(5x^2-3r^2), y(5y^2-3r^2), z(5z^2-3r^2)\right),\\
\label{eq:octupole_3}
(X_x^\beta, X_y^\beta, X_z^\beta) &\propto \left(x(y^2-z^2), y(z^2-x^2), z(x^2-y^2)\right),
\end{align}
where $X$ represents the types of multipoles.
When $X$ corresponds to the time-reversal-odd polar (axial) tensor, it stands for $T$ ($M$) for MT (magnetic) multipole.

By using the multipole notation, the collinear AFM with ${\bm q}={\bm 0}$ on a zigzag chain are represented by the MT dipole $T_z$ when the AFM moment is along the $x$ direction as
\begin{align}
T_z &=\frac{1}{2}\sum_{l={\rm A}, {\rm B}} \left({R}_l^x  \sigma_l^y-{R}_l^y   \sigma_l^x \right)
\to \frac{1}{2} \left(\sigma_{\rm B}^x - \sigma_{\rm A}^x\right),
\end{align}
where $\sigma_l^\mu$ and ${R}_l^\mu$ ($\mu=x,y,z$) are the magnetic moment and the position vector at $l$th atom, respectively~\cite{suzuki2019multipole}.
Similarly, the AFM with the moment along the $y$ direction is characterized by
the magnetic quadrupole $M_u$ as
\begin{align}
M_u &= \sum_{l={\rm A}, {\rm B}} \left(2{R}_l^z  \sigma_l^z -  {R}_l^x  \sigma_l^x - {R}_l^y  \sigma_l^y\right)
\to \frac{1}{2}\left(\sigma_{\rm B}^y - \sigma_{\rm A}^y\right),
\end{align}
and that along the $z$ direction is by the MT dipole $T_x$ as
\begin{align}
T_x &=\frac{1}{2}\sum_{l={\rm A}, {\rm B}} \left({R}_l^y  \sigma_l^z-{R}_l^z   \sigma_l^y \right)
\to \frac{1}{2} \left(\sigma_{\rm A}^{z} - \sigma_{\rm B}^{z} \right).
\end{align}

Moreover, the dipole and octupole components of $\sigma_{\mu\nu\lambda}$ in Eq.~\eqref{eq:rank12} are related to the MT dipoles in Eq.~\eqref{eq:dipole} and MT octupoles in Eqs.~\eqref{eq:octupole_1}--\eqref{eq:octupole_3} as follows:
\begin{align}
T_x &\leftrightarrow  \frac{1}{15}\sum_{\nu=x,y,z} (\sigma_{x\nu\nu}+2\sigma_{\nu\nu x}) \ (\mbox{cyclic}), \\
T_{xyz} &\leftrightarrow \sigma_{xyz}, \\
T_x^\alpha &\leftrightarrow  \frac{1}{10} \left(5\sigma_{xxx}-3\sum_{\nu=x,y,z}\sigma_{x\nu\nu} \right) \ (\mbox{cyclic}), \\
T_x^\beta &\leftrightarrow  \frac{1}{2} \left(\sigma_{xyy}-\sigma_{zzx} \right) \ (\mbox{cyclic}). 
\end{align} 

\section{Essential model parameters in response tensors \label{sec:extract_parameter}}

We show the essential model parameters for the asymmetric band modulation, the longitudinal and transverse nonlinear conductivities, and the linear Hall and magnetoelectric coefficients, by using the systematic analysis method in Refs.~\cite{hayami2020bottom} and \cite{doi:10.7566/JPSJ.91.014701}.
The results are summarized in Table~\ref{tab_params}.

\subsection{Asymmetric band modulation} \label{sec:abm}

First, we give the essential model parameters for the asymmetric band modulation.
Following the method for extracting the essential model parameters in the thermal average of an hermitian operator~\cite{hayami2020bottom,doi:10.7566/JPSJ.91.014701}, we obtain the momentum distribution of the band modulation and its parameter dependences by analytically evaluating the low-order contributions of the following quantity,
\begin{align}
  \Omega^{i} (\bm{k}) = \mathrm{Tr} \left[h^{i+1}(\bm{k})\right].
  \label{eq:omk_h}
\end{align}
Here $h^{i+1}(\bm{k})$ denotes the $(i+1)$-th power of the Hamiltonian matrix at wave vector $\bm{k}$, i.e., $\mathcal{H}$ in Eq.~\eqref{eq:hamiltonian} is represented as $\mathcal{H}=\sum_{\bm k}h({\bm k})$.
The 0th- and 1st-order contributions $\Omega^{0} (\bm{k})$ and $\Omega^{1} (\bm{k})$ are explicitly given by
\begin{widetext}
\begin{align}
   & \Omega^{0} (\bm{k}) = -8\left(t_{2} \cos k_{x}+t_{3} \cos k_{z}\right),                                                                                                   \\
   & \Omega^{1} (\bm{k}) = - 8 \alpha_{1} \tilde{T}_{x}^{\rm MF} \sin k_{x}+ 4\left[\left(\tilde{T}_{x}^{\rm MF}\right)^{2} +\alpha_1^ 2\sin^2 k_{x} + \alpha_2^ 2\sin^2 k_{z}
    +2 t_{1}^{2} (1+\cos k_{x}) +4\left(t_{2} \cos k_{x}+t_{3} \cos k_{z}\right)^2 \right].
  \label{eq:omk_h_2}
\end{align}
\end{widetext}
The odd function of $k_{x}$ appears only in the first term of Eq.~(\ref{eq:omk_h_2}) in the form proportional to $\alpha_{1} \tilde{T}_{x}^{\rm MF}$, which means that the asymmetric band structure is induced by the coupling between the nonzero $\tilde{T}_x^{\rm MF}$ and $\alpha_1$.
It is confirmed at least to the 6th order.
Note that the odd functions of $k_{x}$ included in the higher-order terms in Eq.~(\ref{eq:omk_h}) are always proportional to $\alpha_1 \tilde{T}_x^{\rm MF}$.
Thus, both $\alpha_1$ and $\tilde{T}_x^{\rm MF}$ are the essential model parameters for the asymmetric band structure, and their coupling is also crucial for nonlinear conductivities.

\subsection{Second-order nonlinear conductivity \label{sec:ncon}}

Next, we elucidate the essential model parameters in the longitudinal and transverse nonlinear conductivities.
The essential model parameters in the Drude-type nonlinear conductivities can be extracted by evaluating the following quantity~\cite{doi:10.7566/JPSJ.91.014701},
\begin{align}
  \mathrm{Re} \left[\Gamma^{ijk}_{\mu\nu\lambda}\right] = \sum_{\bm{k}}\mathrm{Re} \left\{\mathrm{Tr} \left[\hat{\varv}_{\mu \bm{k}}
  h^{i}(\bm{k}) \hat{\varv}_{\nu \bm{k}}
  h^{j}(\bm{k}) \hat{\varv}_{\lambda \bm{k}}
  h^{k}(\bm{k})\right]\right\},
  \label{eq:g_ncon}
\end{align}
where $\hat{\varv}_{\mu \bm{k}}$ denotes the $\mu$ component of the velocity operator at $\bm{k}$.

\begin{figure*}[htb!]
  \centering
  \includegraphics[width=180mm]{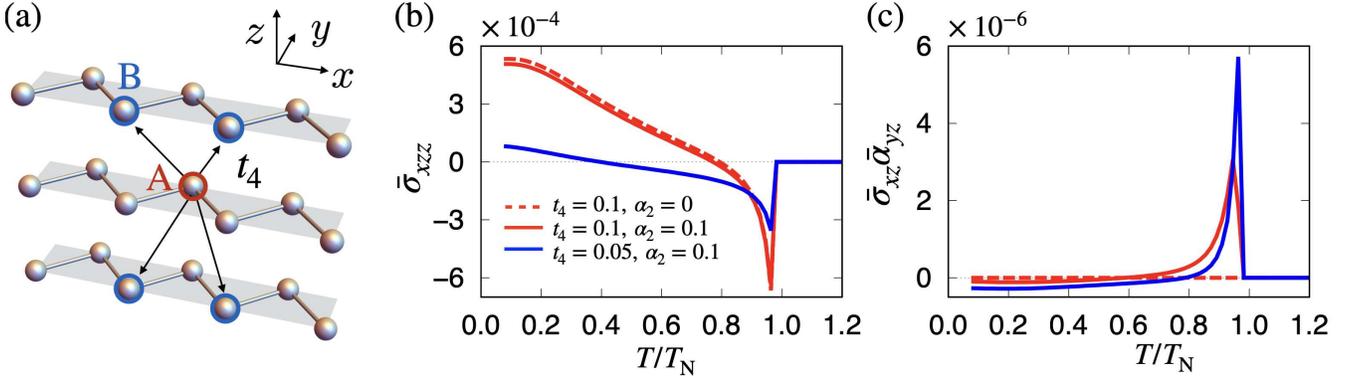}
  \caption{
    (a) Schematic picture of the interlayer hopping $t_4$ between A and B sublattices.
    (b), (c)  The $T$ dependence of (b) $\bar{\sigma}_{xzz}$ and (c) $\bar{\sigma}_{xz}\bar{\alpha}_{yz}$ for $(t_4, \alpha_2)=(0.1,0)$, $(0.1,0.1)$, and $(0.05, 0.1)$.}
    \label{fig:fig_sm}
\end{figure*}

Here, we introduce the interlayer hopping between the sublattices A and B [Fig.~\ref{fig:fig_sm}(a)].
The effect of the additional hopping is taken into account by replacing $\varepsilon^{\rm AB}({\bm k})$ as
$-2t_1\cos{(k_xa/2)} \to -2[t_1+2t_4\cos{(k_zc)}]\cos{(k_xa/2)}$.
The results of the evaluations are given as follows.

\begin{itemize}
  \item[-] \underline{Longitudinal nonlinear conductivity $\sigma_{xxx}$} \\
        As the essential model parameters are included in any pairs of $(i,j,k)$ in Eq.~(\ref{eq:g_ncon}),
        we here show two low-order contributions to Eq.~(\ref{eq:g_ncon}) in the $(i,j,k) = (0,0,1)$ and $(0,1,3)$ terms as representative examples, which are explicitly given by
        \begin{widetext}
        \begin{align}
                       & \mathrm{Re} \left[\Gamma^{001}_{xxx}\right] = \alpha_{1} \tilde{T}_x^{\rm MF} \left(t_{1}^{2}+2 t_{4}^{2}\right), \\
                       & \mathrm{Re} \left[\Gamma^{013}_{xxx}\right] =
          4 \alpha_{1} \tilde{T}_x^{\rm MF}
          \left(
          t_{2} \left\{ \alpha_{1}^{2} \alpha_{2}^{2} + t_{1}^{2} \left[ 4\left(\tilde{T}_x^{\rm MF}\right)^{2} +7 \alpha_{1}^{2} +2 \alpha_{2}^{2}+3 t_{1}^{2} \right]\right\}
          \right. \cr & \hspace{3cm}
            +t_{4} \left[-4\left(\tilde{T}_x^{\rm MF}\right)^{2} t_{1} t_{3}+5 \alpha_{1}^{2} t_{1} t_{3}-\alpha_{2}^{2} t_{1} t_{3}-16 t_{1}^{3} t_{3}-12 t_{1} t_{2}^{2} t_{3}-12 t_{1} t_{3}^{3}
          \right. \cr & \hspace{3cm} \quad
            \left.\left.
            +8\left(\tilde{T}_x^{\rm MF}\right)^{2} t_{2} t_{4} +14 \alpha_{1}^{2} t_{2} t_{4}+2 \alpha_{2}^{2} t_{2} t_{4}+36 t_{1}^{2} t_{2} t_{4}-48 t_{1} t_{3} t_{4}^{3}+18 t_{2} t_{4}^{3}
            \right]\right).
            \label{eq:para_xxx}
        \end{align}
        Then, the essential model parameters in the longitudinal nonlinear conductivity $\sigma_{xxx}$ are $\alpha_1$ and $\tilde{T}_{x}^{\rm MF}$, which is consistent with the fact that the nonzero $\sigma_{xxx}$ is closely related to the asymmetric band structure under $T_x^{\rm MF} \neq 0$.
        Since all the terms in Eq.~(\ref{eq:g_ncon}) are always proportional to $\alpha_1 \tilde{T}_{x}^{\rm MF}$, $\sigma_{xxx}$ is written in the form:
        \begin{align}
          \sigma_{xxx} = \alpha_{1} \tilde{T}_{x}^{\mathrm{MF}}
          \left[
          t_{1}^{2} F(t_1,t_2,t_3,t_4,\alpha_1,\alpha_2,\tilde{T}_x^{\rm MF})
          +t_{2} F'(t_1,t_2,t_3,t_4,\alpha_1,\alpha_2,\tilde{T}_x^{\rm MF})
          + t_{4} F''(t_1,t_2,t_3,t_4,\alpha_1,\alpha_2,\tilde{T}_x^{\rm MF})
          \right],
          \label{eq:ncon_xxx}
        \end{align}
        \end{widetext}
        where the even power of $\alpha_1$ and $\tilde{T}_x^{\rm AF}$ appears in $F$, $F'$, and $F''$, e.g., $\alpha_1^2$ and $(\tilde{T}_x^{\rm MF})^2$ in Eq.~\eqref{eq:para_xxx}.
        By introducing $t_{4} \neq 0$, the additional contribution appears, which results in the alternative behavior of $\sigma_{xxx}$.
  \item[-] \underline{Transverse nonlinear conductivity $\sigma_{xzz}$} \\
        Similar to $\sigma_{xxx}$, we show two low-order contributions to Eq.~(\ref{eq:g_ncon}) in
         the $(i,j,k)=(0,1,0)$ and $(0,1,1)$ terms for example.
         The expressions are given by
        \begin{align}
        &\mathrm{Re}\left[\Gamma^{010}_{xzz}\right] = -\frac{242}{25} \alpha_{1} \tilde{T}_x^{\rm MF}  t_{4}^{2}, \\
         & \mathrm{Re} \left[\Gamma^{011}_{xzz}\right] = \frac{121}{25} \alpha_{1} \tilde{T}_{x}^{\rm MF}\left[\alpha_{2}^{2} t_{2}+ t_{4} \left( 4 t_{1} t_{3}+8 t_{2} t_{4}\right)\right].
        \end{align}
        Similar to this result, we find that all the terms in Eq.~(\ref{eq:g_ncon}) are always proportional to $\alpha_{1} \tilde{T}_{x}^{\rm MF}$, then $\sigma_{xzz}$ is expressed as
        \begin{align}
          \sigma_{xzz} =&
          \alpha_1\tilde{T}_x^{\rm MF} \left[\alpha_2^2t_2F(t_1,t_2,t_3,t_4,\alpha_1,\alpha_2,\tilde{T}_x^{\rm MF})\right.\notag\\
          &\left.\qquad \quad+t_4F'(t_1,t_2,t_3,t_4,\alpha_1,\alpha_2,\tilde{T}_x^{\rm MF})\right],
          \label{eq:ncon_xzz}
        \end{align}
        where the second term proportional to $t_4$ does not vanish even for $\alpha_2=0$.
\end{itemize}

\subsection{Linear responses \label{sec:extract_parameter_linear}}

We further clarify the essential model parameters for the linear Hall and magnetoelectric coefficients.
The essential model parameters in the inter-band contribution of the electric-field induced response tensors can be extracted by evaluating the following quantity~\cite{doi:10.7566/JPSJ.91.014701},
\begin{align}
  \mathrm{Im}\left[\Gamma^{ij}_{\mu\nu}\right] = \sum_{\bm{k}}\mathrm{Im}\left\{\mathrm{Tr} \left[\hat{A}_{\mu \bm{k}}
  h^{i}(\bm{k}) \hat{\varv}_{\nu \bm{k}}
  h^{j}(\bm{k})\right]\right\},
  \label{eq:g_le}
\end{align}
where $\hat{A}_{\mu \bm{k}}$ denotes the $\mu$ component of an arbitrary hermitian operator at $\bm{k}$.

\begin{itemize}
  \item[-] \underline{Magnetoelectric coefficient $\alpha_{yz}$} \\
        The magnetoelectric coefficient $\alpha_{yz}$ corresponds to the case with $\hat{A}_{\mu \bm{k}}=\sigma_{y}$ in Eq.~(\ref{eq:g_le}).
        Similar to the nonlinear conductivities, the essential model parameters are included in any pairs of $(i,j)$ in Eq.~(\ref{eq:g_le}).
        We show two cases by taking $(i,j)=(0,2)$ and $(1,3)$, which are given by
        \begin{widetext}
        \begin{align}
        &\mathrm{Im}\left[\Gamma_{yz}^{02}\right] = -\frac{44}{5}  \alpha_{2}\tilde{T}_x^{\rm MF} t_{3},\\
         &\mathrm{Im}\left[\Gamma^{13}_{yz}\right]
          = \frac{11}{5} \alpha_{2} \tilde{T}_x^{\rm MF} \left\{t_{3}\left[ 4\left(\tilde{T}_x^{\rm MF}\right)^{2} +6 \alpha_{1}^{2} +\alpha_{2}^{2} +8 t_{1}^{2} -24 t_{2}^{2} -12 t_{2}\right] +t_{4} \left(16 t_{1} t_{2} +24 t_{3} t_{4}\right) \right\}.
        \end{align}
	\end{widetext}
        We also find that all the terms in Eq.~(\ref{eq:g_le}) are always proportional to $\alpha_{2} \tilde{T}_x^{\rm MF}$, then $\alpha_{yz}$ is expressed as
        \begin{align}
          \alpha_{yz} =& \alpha_{2} \tilde{T}_x^{\rm MF} \left[t_3 F(t_1,t_2,t_3,t_4,\alpha_1,\alpha_2,\tilde{T}_x^{\rm MF})\right.\notag\\
          &\left.\qquad\quad +t_4 F'(t_1,t_2,t_3,t_4,\alpha_1,\alpha_2,\tilde{T}_x^{\rm MF})\right].
          \label{eq:a_yz}
        \end{align}
        Therefore, the essential model parameters are $\alpha_{2}$
        and $\tilde{T}_x^{\rm MF}$, while $\alpha_{yz}$ also depends on the spin-independent hopping process $t_{3}$ or $t_{4}$.
  \item[-] \underline{Hall coefficient $\sigma_{xz}$} \\
        In order to discuss $\sigma_{xz}$, we introduce the small magnetic field along the $y$ direction $H_y$.
        Then, we evaluate the essential model parameters for the Hall coefficient $\sigma_{xz}$ with $\hat{A}_{\mu \bm{k}}=\hat{\varv}_{x \bm{k}}$ in Eq.~(\ref{eq:g_le}).
        We show two low-order contributions to Eq.~(\ref{eq:g_le}) in the $(i,j)=(0,3)$ and $(1,3)$ terms for example, which are given by
        \begin{align}
           &
            \mathrm{Im}\left[\Gamma^{03}_{xz}\right] = \frac{44}{5} \alpha_{1} \alpha_{2} H_{y} \left(3 t_{2} t_{3} + 5 t_{1} t_{4}\right),\\
           & \mathrm{Im}\left[\Gamma^{13}_{xz}\right]
          =
          \frac{88}{5} \alpha_{1} \alpha_{2} H_{y} \left[2 t_{1}^{2} t_{3}+ t_{4} \left( 8 t_{1} t_{2}+7 t_{3} t_{4}\right) \right].
        \end{align}
        All the terms in Eq.~(\ref{eq:g_le}) are always proportional to $\alpha_{1} \alpha_{2} H_{y}$, then $\sigma_{xz}$ is expressed as
        \begin{align}
          \sigma_{xz} =& \alpha_{1} \alpha_{2} H_{y} \left[t_3 F(t_1,t_2,t_3,t_4,\alpha_1,\alpha_2, H_y,\tilde{T}_x^{\rm MF})\right.\notag\\
          &\left. \qquad \quad+t_4 F'(t_1,t_2,t_3,t_4,\alpha_1,\alpha_2, H_y,\tilde{T}_x^{\rm MF})\right].
          \label{eq:lcon_xz}
        \end{align}
\end{itemize}
Therefore, the essential model parameters are $\alpha_{1}$, $\alpha_{2}$, and $H_y$, while $\sigma_{xz}$ also depends on the spin-independent hopping along the $z$ direction, $t_{3}$ or $t_{4}$.

By combining the results, Eqs.~(\ref{eq:a_yz}) and (\ref{eq:lcon_xz}), $\sigma_{xz}\alpha_{yz}$ has the form:
\begin{widetext}
\begin{align}
  \sigma_{xz}\alpha_{yz} = \alpha_{1} \alpha_{2}^{2} \tilde{T}_{x}^{\rm MF} H_{y} \left[
  t_{3}^{2} F(t_1,t_2,t_3,t_4,\alpha_1,\alpha_2, H_y,\tilde{T}_x^{\rm MF})
  + t_{4}^{2} F'(t_1,t_2,t_3,t_4,\alpha_1,\alpha_2, H_y,\tilde{T}_x^{\rm MF})
  + t_{3}t_{4} F''(t_1,t_2,t_3,t_4,\alpha_1,\alpha_2, H_y,\tilde{T}_x^{\rm MF})
  \right],
  \label{eq:alcon}
\end{align}
\end{widetext}
which clearly shows that $\sigma_{xz}\alpha_{yz} \propto \alpha_{1} \alpha_{2}^{2} \tilde{T}_{x}^{\rm MF} H_{y}$ irrespective of the additional parameter of $t_{4}$.

When $t_{4} = 0$, we find that both $\sigma_{xzz}$ and $\sigma_{xz}\alpha_{yz}$ are proportional to $\alpha_1 \alpha_2^2 \tilde{T}_{x}^{\rm MF}$.
On the other hand, such relation does not hold when $t_{4} \neq 0$; $\sigma_{xzz} \propto \alpha_1 \tilde{T}_{x}^{\rm MF}$, whereas $\sigma_{xz}\alpha_{yz} \propto \alpha_1 \alpha_2^2 \tilde{T}_{x}^{\rm MF}$.

\section{Effect of additional interlayer hopping \label{sec:interlayer_hopping}}

We compare the transverse component of the nonlinear conductivity $\sigma_{xzz}$ and the quantity $\sigma_{xz}\alpha_{yz}$
in the presence of the interlayer hopping $t_{4}$ between the sublattices A and B.

Figures~\ref{fig:fig_sm}(b) and \ref{fig:fig_sm}(c) show $\bar{\sigma}_{xzz}$ and $\bar{\sigma}_{xz}\bar{\alpha}_{yz}$ as functions of $T$, respectively, for  $t_4=0.1,0.05$ and $\alpha_2=0,0.1$, where $\alpha_1=0.4$ is used.
As shown by the red dashed line in Fig.~\ref{fig:fig_sm}(b), $\bar{\sigma}_{xzz}$ still remains nonzero even for $\alpha_2=0$, while $\bar{\sigma}_{xz}\bar{\alpha}_{yz}$ in Fig.~\ref{fig:fig_sm}(c) vanishes.
Furthermore, the nonzero $t_4$ enhances $\bar{\sigma}_{xzz}$, while it suppresses $\bar{\sigma}_{xz}\bar{\alpha}_{yz}$ while increasing $t_4$.
This is because the essential model parameters discussed in the previous section are different for $\sigma_{xzz}$ and $\sigma_{xz}\alpha_{yz}$.
Indeed, in the presence of $t_4$ and $\alpha_2$, the essential model parameter of $\sigma_{xzz}$ is represented as
$\alpha_1\tilde{T}_x^{\rm MF}[\alpha_2^2t_2F(t_1,t_2,t_3,t_4,\alpha_1,\alpha_2,\tilde{T}_x^{\rm MF})+t_4F'(t_1,t_2,t_3,t_4,\alpha_1,\alpha_2,\tilde{T}_x^{\rm MF})]$,
which clearly shows that $\sigma_{xzz}$ has the additional contribution from $t_4$ and does not vanish for $\alpha_2=0$.
On the other hand, the essential model parameters of $\sigma_{xz}$ and $\alpha_{yz}$ does not show the change from
$\sigma_{xz} \to \alpha_1\alpha_2H_y F(t_1,t_2,t_3,t_4,\alpha_1,\alpha_2,H_y,\tilde{T}_x^{\rm MF})$ and $\alpha_{yz} \to \alpha_2\tilde{T}_x^{\rm MF} F(t_1,t_2,t_3,t_4,\alpha_1,\alpha_2,\tilde{T}_x^{\rm MF})$, respectively; the hopping $t_4$ is not the essential model parameter for $\sigma_{xz}$ and $\alpha_{yz}$.
Thus, there is no simple relation between them.

\bibliographystyle{apsrev4-2}
\bibliography{ref}
\end{document}